**End-Stage Liver Disease Comorbidities in Patients Awaiting Transplantation: Identification and Impact on Liver Transplant Survival**

Julia Tacherel,[1] Kiruthika Balakrishnan,[1] Gyorgy Simon,[2] Lisiane Pruinelli[1,3]

[1] College of Nursing, University of Florida, Gainesville, FL

[2] School of Medicine and Institute for Health Informatics, University of Minnesota, Minneapolis, MN

[3] College of Medicine, University of Florida, Gainesville, FL

*Correspondence*: Lisiane Pruinelli, lisianepruinelli@ufl.edu

**Background**: End Stage Liver Disease (ESLD) is a complex disease process with a high incidence of co-occurring conditions affecting all body systems, negatively impacting overall health. There is a lack of clear understanding and evidence-based practice (EBP) guidelines of the occurrence and progression of common comorbidities, specifically developed for patients awaiting liver transplantation (LT) and its impact on transplant survival and quality of life pre- and post-transplant.

**Aim**: To enhance clinician knowledge in identifying and quantifying the trajectory of the most common and deteriorating comorbidities in ESLD patients awaiting LT and to analyze their impact on patient outcomes.

**Methods**: This is a mixed methods study design. The exploratory phase determined the most frequent ESLD conditions. For each, a review was conducted to extract relevant, EBP-driven data in diagnosing and measuring their severity progression. The data was clustered into five respective research matrices. Using the matrices, the quantitative phase, used a retrospective study design using longitudinal de-identified data from electronic health records (EHR). Data included demographics, laboratory, procedures and medications for those who underwent LT between 2011 and 2021 to determine the presence and progression of each comorbidity. Descriptive statistics and a survival analysis with a Cox Proportional Hazard Model were used to analyze whether comorbidities were associated with transplant survival.

**Results**: The five most frequent included comorbidities were Diabetes Mellitus (DM), Chronic Kidney Disease (CKD), Malnutrition, Portal Hypertension (PH), and Ascites. From the 722 patients, 68.2% were males with a mean age of 54.81 years (sd=11.24) and 19.8% of patients died post-LT. In the probability of surviving over time, survival gradually decreases, with the most notable decline observed at 5 years post-LT. The Cox Proportional Hazard Model showed that age at transplant (p=0.01), waitlist time (p=0.004), DM at listing (p=0.02), and low albumin (p=0.03) and CKD 5 (p=0.04) development after listing are statistically significant predictors of post-LT survival.

**Conclusion**: Many ESLD comorbidities have provider variability in diagnosing and measuring progression. This study adds to this unknown body of knowledge and specifically identifies metrics and quantifies comorbidities progression for ESLD patients impacting post-LT survival. Further studies should investigate how to elucidate the development and trajectories of ESLD comorbidities to derive ESLD specific EBP guidelines that can provide better and personalized care for patients experiencing these conditions.

## Introduction

In the United States (U.S.), over 5 million adults suffer from chronic liver disease (LD) and the majority will progress into End-Stage Liver Disease (ESLD) (Hansen et al., 2004).[1] This patient population experiences a high burden of system-wide illness as progression into irreversible liver damage occurs in a cascading, or exponential, effect. In such cases, liver transplantation (LT) becomes the sole hope for survival. The U.S. stands as a global leader in LT procedures annually (>10,000 in 2023, costing >8 billion),[2] but does not approach liver organ distribution according to a whole-health perspective.[3,4] Current practice follows an urgency model established by the Organ Procurement and Transplantation Network (OPTN), known as the Model for End-Stage Liver Disease (MELD),[5,6] where the sickest patients are prioritized for transplant. The MELD model is a scoring system based on objective laboratory data findings and calculates a score that captures the patient's liver function at a specific time. MELD scores fail to provide information regarding future disease progression, as it only predicts the risk of mortality over 90-days while on the waitlist, and do not account for overall health status beyond liver function. Considering these limitations, at least 20% of ESLD patients are delisted for becoming too sick to undergo LT or die while waiting for a liver organ with post-LT survival outcomes ranging from 55.6% to 86.5%.[2] Comprehensively, understanding ESLD disease trajectory and the impact of body-wide comorbidities on patient deterioration, and consequently, mortality, is imperative to accurately predict LT outcomes and fairly care for patients waiting for transplantation.

In a previous retrospective cohort study, comorbidities and risk factors present in a cohort of 344 adult ESLD patients were clustered, or grouped, by body system and tested for impact on post-LT mortality, resulting in three significant clusters. Namely, the presence of conditions affecting the circulatory, endocrine, and musculoskeletal systems was highly predictive of post-LT mortality by the 5-year mark.[3] While the study used a small and single center sample, it demonstrated that comorbidities involving all body systems should not be considered as isolated conditions in ESLD, rather, they have a summative effect on LT outcomes, and be incorporated into evidence-based guidelines.

Evidence-based practice (EBP) guidelines[7,8] are meticulously crafted protocols designed to standardize clinical practice and optimize patient outcomes by integrating the best available research evidence with clinical expertise. These guidelines are generally developed through comprehensive reviews of current literature, expert consensus, and analysis of population-level data, ensuring they address the needs of the general population. However, this broad approach often results in a lack of specificity for particular subpopulations, such as patients awaiting liver transplants, who present unique clinical challenges and variability. This shortfall underscores the critical need for developing personalized guidelines that take into account the specific characteristics, comorbidities, and treatment responses of these specialized groups. The development and adoption of such personalized guidelines would ensure more accurate and effective care, ultimately improving outcomes for waitlisted liver transplant patients.

Therefore, this study aims to map out the most frequent data-driven identified comorbidities common in ESLD patient populations to current EBP guidelines, and to understand the impact of having or developing new comorbidities on LT mortality. Meaningful results will prompt targeted diagnostics, interventions, and future research; moving towards personalized EBP, holistic patient care.

**Methods**

This is a mixed methods study design. The exploratory phase determined the most frequent ESLD conditions, as determined by the study cohort. For each, a literature review was conducted to extract relevant, evidence-based practice (EBP) data in diagnosing and measuring their severity progression. The second phase included a retrospective study design using longitudinal de-identified data from electronic health records (EHR) available through the University of Minnesota Clinical and Translational Science Institute (CTSI).[9] All data were managed and analyzed inside the CTSI secure environment and were determined as exempt from involving human subjects by the Institutional Review Board (#00000092).

**Cohort Selection**

The cohort included all adult patients (18 years or older at the time of LT) who underwent LT for the first time between 2011 and 2021. Patients were excluded if they had missing data to establish the correct LT date, combined transplantation (i.e., liver along with lungs, pancreas, and/or heart), and if missing organ or end-point information. Patients who had combined liver and kidney were not excluded as kidney transplant is common and is used as a therapeutic requirement in many LT cases. Demographic information, laboratory values, encounter information, comorbidities, and the outcomes of interest (mortality) data for all patients were extracted.

**ESLD EBP Guideline Matrices**

First, the exploratory phase determined the most frequent ESLD conditions. We focused on identifying comorbidities that result both from liver decompensation over time and have the potential to be prevented if patients are intervened timely. These comorbidities were identified based on prevalence in the included cohort and current data on the ESLD patient population progression. The most frequent five ESLD comorbidities identified were Diabetes Mellitus (DM), Chronic Kidney Disease (CKD), Malnutrition, Portal Hypertension (PH), and Ascites.

ESLD EBP guideline data were extracted using current best EBP guidelines for each comorbidity from the literature.

The ESLD EBP guideline matrices (Appendix A) were organized into a similar sequence with four main sections: an overarching description of the condition, diagnostic and staging measures, severity measures, and complications based on acuity. The diagnostic and staging section further breaks down into the main diagnostic requirements followed by varying sub-sections that provide specific detail and comments. All components included in the matrices, including objective or numerical findings and otherwise, provide various measures of tracking disease progression and deterioration. In terms of formatting, the information flows vertically between the main sections and horizontally between the sub-sections. Each matrix was used to guide the identification of comorbidities and related deterioration within the EHRs of the included cohort.

**Included Features**

Demographic data included age at LT, sex, and waitlist time. Additionally, MELD scores at listing and at delisting (i.e., at the time of LT) were included, and Delta MELD was calculated as the difference between MELD these two values. Laboratory, medications, procedures, problems and diagnosis were used to build the features. Comorbidities that occurred before LT listing were considered as present at the LT listing date, and comorbidities that developed throughout the waitlist time and up until one day before LT date were considered a new comorbidity. The decision to include up to one day before LT was to prevent from capturing any problem or comorbidity resulting from the LT procedure. The outcome of interest was recipient overall survival over time post-LT. Patients were censored at the last follow-up and considered alive or dead at that point.

**Statistical Analysis**

Continuous features are described with mean and standard deviations (sd) when normally distributed and as median and interquartile range (IQR) for skewed distributed features. Categorical variables are described as counts (n) and percentages (%). Mortality, age, and waitlist time were treated as normally distributed. MELD at listing was treated as a skewed distribution. A sensitivity analysis was performed to evaluate if there was a statistically significant difference between who was alive or not at the last follow-up appointment. Survival tables were used to analyze the survival probabilities over time. A Cox proportional hazard model with backward elimination was further used to model the relationship between the risk factors and the outcome of interest. Statistically significant variables were considered at alpha $<0.05$.

**Results**

**ESLD EBP Guideline Matrices**

Data extraction on the five comorbidities were primarily taken from five recognized sources including The American Diabetes Association (ADA),[10] Kidney Disease Improving Global Outcomes (KDIGO),[11] The European Society of Clinical Nutrition and Metabolism (ESPEN),[12] The Global Leadership Initiative on Malnutrition (GLIM),[13] and The American Association for the Study of Liver Diseases (AASLD).[14,15] Each EBP diagnostic guideline criteria were the most current and up-to-date information with the exception of CKD. The KDIGO Clinical Practice Guideline for the Evaluation and Management of Chronic Kidney Disease from the official *Journal of the International Society of Nephrology* was based on the 2012 guidelines (2013), and an updated (2024) guideline was published after this study had performed data collection.

The resulting ESLD EBP matrix (Appendix A) included DM which was further separated into prediabetes, Type 1 Diabetes Mellitus (T1DM), and Type 2 Diabetes Mellitus (T2DM). While the information is very similar between T1DM and T2DM, slight differences that add context for measuring the progression of illness should be noted. Namely, in T1DM, deterioration of illness can be measured with increased insulin needs for blood glucose management, whereas in T2DM, deterioration can be also measured with an increase from oral anti-diabetic medications to insulin. In the matrix outlining CKD, staging is based on different combinations of eGFR and Albuminuria categories. An acute kidney injury (AKI) can both precede CKD and be an acute complication of it, therefore serves as a good marker for measuring kidney deterioration throughout the progression of illness. The matrix for Malnutrition, which is multifactorial and, thus, diagnosis can be highly variable and patient dependent. In other words, while this matrix is a strong guide for the diagnosis and staging of malnutrition, it also requires a strong foundation of clinical judgment for accuracy. As for the matrix outlining PH, there are many combinations of invasive and non-invasive techniques and tools for diagnosis. This is helpful when patient data and the availability of testing is limited. Particularly, catheterization of the hepatic vein is the gold standard for diagnosing PH, but it is an expensive and invasive procedure that has limited availability in many clinical settings, and not all clinicians document the amount of pressure identified. Finally, the matrix for Ascites is clearer in its diagnosis. Regardless of the tool used, confirming any amount of peritoneal fluid in the abdomen confirms the diagnosis of Ascites. The difference between the use of tools is important when measuring the severity of illness. Requiring a non-invasive removal of fluid with the use of medications such as diuretics versus requiring an invasive removal of fluid by abdominal paracentesis is an important severity measure.

## Cohort Description

Out of the initially identified 826 patients who underwent LT, 14 patients were excluded because the LT date was before 2011, 88 patients were younger than 18 years old at the time of LT, and 1 patient had a combined transplant, totaling 722 patients for inclusion (Table 1). The final cohort had a mean age of 54.81 years (sd=11.24), 68.2% of patients were males, and 19.8% of patients died post-LT. Significant differences between the two groups of survived and not survived at the last follow up, are noted in several variables, suggesting potential factors associated with survival after LT.

**Table 1.** Descriptive Statistics of the included cohort (n=722).

| **Characteristic** | **Survived (n=579) (mean±sd/median [IQR]/count(%))** | **Not-Survived (n=143) (mean±sd/median [IQR]/count(%))** | **p-value*** |
|---|---|---|---|
| Age at transplant (Years)<br><45<br>>=45 | 54.06 ± 11.62<br><br>105 (18.13)<br>474 (81.86) | 57.83 ± 8.99<br><br>12 (8.39)<br>131 (91.60) | <0.001<br>0.006 |
| Sex (Male) | 101 (17.44) | 42 (29.37) | 0.57 |
| Waitlist time (Days) | 138 [15-361] | 232 [30-495] | 0.001 |
| MELD at listing | 20 [12.5-29] | 16 [10-28] | 0.008 |
| MELD at delisting | 23 [13-33] | 22 [12-34] | 0.95 |
| MELD by age group (<45) | 22 [13-33] | 28 [20-35] | <.001 |
| Delta MELD | 0 [-1-3] | 1 [0-7] | <0.001 |
| Simultaneous liver-kidney transplant | 29 (5.00) | 13 (9.09) | 0.09 |
| Hepatocarcinoma | 165 (28.49) | 46 (32.16) | 0.44 |
| Albumin at listing | 2.93 ± 0.61 | 3.08 ± 0.65 | 0.06 |
| Waitlist albumin <3.4 | 30 (5.18) | 20 (13.98) | <.001 |
| Proteinuria at listing | 34.74 ± 102.26 | 50.06 ± 97.38 | 0.18 |
| Median Follow-up (Days) | 1511 [616-2478] | 803 [137-1621] | <0.001 |
| **Diabetes Mellitus (DM)** | | | |
| At listing<br>Pre-diabetes<br>Diabetes<br>Insulin use<br>Oral antidiabetic medication | <br><11 (<1.89)<br>132 (22.80)<br>81 (13.98)<br>106 (18.30) | <br>0 (0)<br>40 (27.97)<br>23 (16.08)<br>29 (20.27) | <br>1<br>0.23<br>0.61<br>0.29 |
| Post listing<br>Pre-diabetes<br>Diabetes<br>Insulin use<br>Oral antidiabetic medication | <br>0 (0)<br>64 (11.05)<br>66 (13.77)<br>68 (11.74) | <br>0 (0)<br>24 (16.78)<br>27 (18.88)<br>27 (18.88) | <br>NA<br>0.08<br>0.02<br>0.29 |
| **Chronic Kidney Disease (CKD)** | | | |
| At listing<br>Stage 1<br>Stage 2<br>Stage 3a<br>Stage 3b<br>Stage 4<br>Stage 5 | <br>157 (27.46)<br>301 (51.98)<br>40 (6.90)<br>31 (5.35)<br>29 (5.00)<br>21 (3.62) | <br>29 20.27)<br>83 (58.04)<br>12 (8.39)<br><11 (<7.69)<br><11 (<7.69)<br><11 (<7.69) | <br>1<br>1<br>1<br>1<br>1<br>1 |
| Post listing<br>Stage 1<br>Stage 2<br>Stage 3a<br>Stage 3b<br>Stage 4<br>Stage 5 | <br>96 (16.58)<br>293 (50.60)<br>53 9.15)<br>58 (10.01)<br>47 (8.11)<br>32 (5.52) | <br><11 (<7.69)<br>75 (52.44)<br>18 (12.58)<br><11 (<7.69)<br>17 (11.88)<br>14 (9.79) | <br>0.005<br>0.76<br>1<br>0.25<br>0.20<br>0.09 |
| **Malnutrition** | | | |
| At listing | 18 (3.10) | <11 (<7.69) | 0.01 |
| Post-listing | 11 (1.89) | <11 (<7.69) | 0.95 |
| Dietary Supplements | 93 (16.06) | 29 (20.27) | 0.27 |
| **Anemia** | | | |
| At listing | 134 (23.14) | 25 (17.48) | 0.17 |
| Post-listing | 50 (8.63) | 12 (8.39) | 1 |
| **Portal Hypertension (PH)** | | | |
| At listing | 127 (21.93) | 16 (11.19) | 0.006 |
| Post-listing | 30 (5.18) | 17 (11.89) | 0.006 |
| **Ascites** | | | |
| At listing | 317 (54.74) | 54 (37.76) | <0.001 |
| Post-listing | 58 (10.01) | 24 (16.78) | 0.03 |

Age at transplant shows a significant difference, with the mean age being lower in the survived group (54.06 years) compared to the not-survived group (57.83 years), suggesting that younger age may be associated with better survival outcomes ($p < 0.001$). The median waitlist time and follow-up post-LT also show significant differences; patients who survived had shorter waitlist times (138 days) and longer follow-up periods (1511 days) compared to those who did not survive (both with $p < 0.001$). These findings might indicate that shorter wait times and longer post-transplant monitoring are beneficial.

Certain health metrics at the listing, such as MELD score, malnutrition, PH, and ascites were statistically significant different between the two groups. Although MELD at delisting didn't show statistically significant difference, the Delta MELD showed statistically significance, indicating that the amount of change in liver markers between listing and delisting, as captured by MELD, is significant different between who survived and not-survived. Lower MELD scores in the surviving group suggest that patients with better liver function at the time of listing have improved survival rates. The presence of PH was also more common in the not-survived group, highlighting its potential impact on survival outcomes (p= 0.006). In contrast, the presence of DM showed no significant difference at listing. However, starting of insulin use post-listing was statistically significant between the two groups, showing a potential impact on the severity of DM on post-LT survival, when DM further progress while on the waitlist.

Progression into CKD stage 5, particularly post-listing and having CKD 5 at the time of LT, showed statistically significant differences (p=0.09), pointing to the importance of aggressive and ongoing medical management of kidney organ health to prevent progression throughout the LT to improve survival probabilities. Both at-listing and post-listing presence of ascites had a statistically significant difference between survived and not survived groups (p-= <0.001 and p=0.03, respectively). At listing, a considerably higher percentage of patients who did not survive had ascites compared to those who did, suggesting that ascites is associated with worse survival outcomes. This pattern is consistent with post-listing, reinforcing ascites as a critical marker of liver disease severity affecting patient prognosis post-transplant.

**Risk Factors Associated with Survival**

The final Cox Proportional Hazard Model (Table 2) outlines how various factors are associated with the risk, or hazard, of patient death post-LT. Five significant risk factors increase the likelihood of mortality post-LT, such as age at transplant (p0.01), waitlist time (p=0.004), DM at listing (p=0.02), developing low albumin after listing (p=0.03), and development Ing CKD5 after listing (p= 0.04). Particularly, for every additional year of age at the time of transplant, the hazard increases by 1.02%, therefore the older patients become the higher is the likelihood of dying post-LT. Similarly, the longer time patients spend waiting for LT there is an increase in the likelihood of mortality post-LT. Finally, developing low albumin and CKD stage 5 throughout the waitlist period increases the hazard by 75% and 75%%, respectively, indicating a considerably higher likelihood for death post-LT associated with worsened of these two conditions.

The other factors were not statistically significant on post-LT mortality, although a few key considerations regarding PH. Firstly, the negative coefficient for PH at listing suggests that having PH at listing might be associated with a lower risk of death post-LT. This indicates that this comorbidity, considering the timing of their development, may have a protective effect on patients. This protective effect may be due to additional therapies and treatments received when these comorbidities were diagnosed at the first place; however, without analyzing the different therapeutic strategies patients received as well as their overall state of health at that time, it is hard to draw any further conclusions. In all, significant results should be considered when listing patients for LT and managing them throughout the process, and non-significant results should be considered when conducting future research on the ESLD population awaiting LT.

**Table 2**

*Cox Proportional Hazard Model*

| | **Coefficient** | **HR** | **95% CI** | ***se*** | ***p*-value** |
|---|---|---|---|---|---|
| **Age at transplant (years)** | 0.0206 | 1.02 | [1.00-1.03] | 0.0088 | 0.01 |
| **Waitlist time (Days)** | 0.0003 | 1.00 | [1.00-1.00] | 0.0001 | 0.004 |
| **Delta MELD** | 0.0219 | 1.02 | [0.99-1.04] | 0.0118 | 0.06 |
| **Diabetes Mellitus at Listing** | 0.4483 | 1.56 | [1.06-2.29] | 0.1957 | 0.02 |
| **Portal Hypertension at Listing** | -0.4906 | 0.61 | [0.36-1.04] | 0.2709 | 0.07 |
| **Low Albumin after listing** | 0.5501 | 1.73 | [1.04-2.87] | 0.2574 | 0.03 |
| **Chronic Kidney Disease Stage 5 at LT** | 0.5616 | 1.75 | [1.00-3.05] | 0.2832 | 0.04 |

**Survival Overtime**

When measuring the probability of surviving over time (Table 3), data was measured in specific time intervals from 30 days post-LT up to 1,825 days, or 5 years, post-LT. Results show that survival probability over time decreases as expected in cohort studies involving patients with progressive diseases and a higher risk of mortality, especially following transplantation. Overall, the cohort initially exhibits a high survival probability that gradually decreases over time, with the most notable decline observed at 5 years post-LT.

**Table 3**

*Probability of Surviving Overtime for the included cohort after LT (n=722)*

| **Time (Days)** | **Subjects at Risk** | **Number of Events** | **Survival Probability** | **Standard Error** | **95% Lower CI** | **95% Upper CI** |
|---|---|---|---|---|---|---|
| 30 | 700 | 18 | 0.975 | 0.006 | 0.964 | 0.987 |
| 90 | 665 | 16 | 0.953 | 0.008 | 0.937 | 0.968 |
| 365 | 570 | 22 | 0.919 | 0.01 | 0.898 | 0.939 |
| 730 | 484 | 13 | 0.896 | 0.012 | 0.873 | 0.920 |
| 1825 | 268 | 42 | 0.800 | 0.018 | 0.766 | 0.835 |

During the exploratory analysis, data shows that the mean age has been changing in recent years with more young patients being added to the waitlist, then what was reported in previous cohort studies, specifically for those patients who survived. Further exploration into age differences showed that younger patients (<45 years old) are sicker, as measured by MELD at the time of LT, when compared to >=45 years old patients. When evaluating the relationship between age groups and MELD, results showed, although weak, a statistically significant negative relationship ($r = -0.12$, $p = <.001$), showing that >=45 years old patients undergoing liver transplant have a decreased MELD at the time of transplant than <45 years old patients (Figure 1). Specifically, patients younger than 45 years old had a statistically significant higher MELD than any other age groups at the time of transplant ($p < 0.001$). However, even considering their statistically significant higher MELD before LT, patients younger than 45 years old, had statistically significant higher survival probabilities than any other age groups ($p = 0.031$) (Figure 2), reinforcing previous findings that MELD is not able to estimate post-LT survival and other factors are better determinants of patients' survival over time. We further looked into patients who had Hepatocarcinoma and those who didn't, and results showed no statistically significant difference ($p = 0.32$) in survival between the two groups.

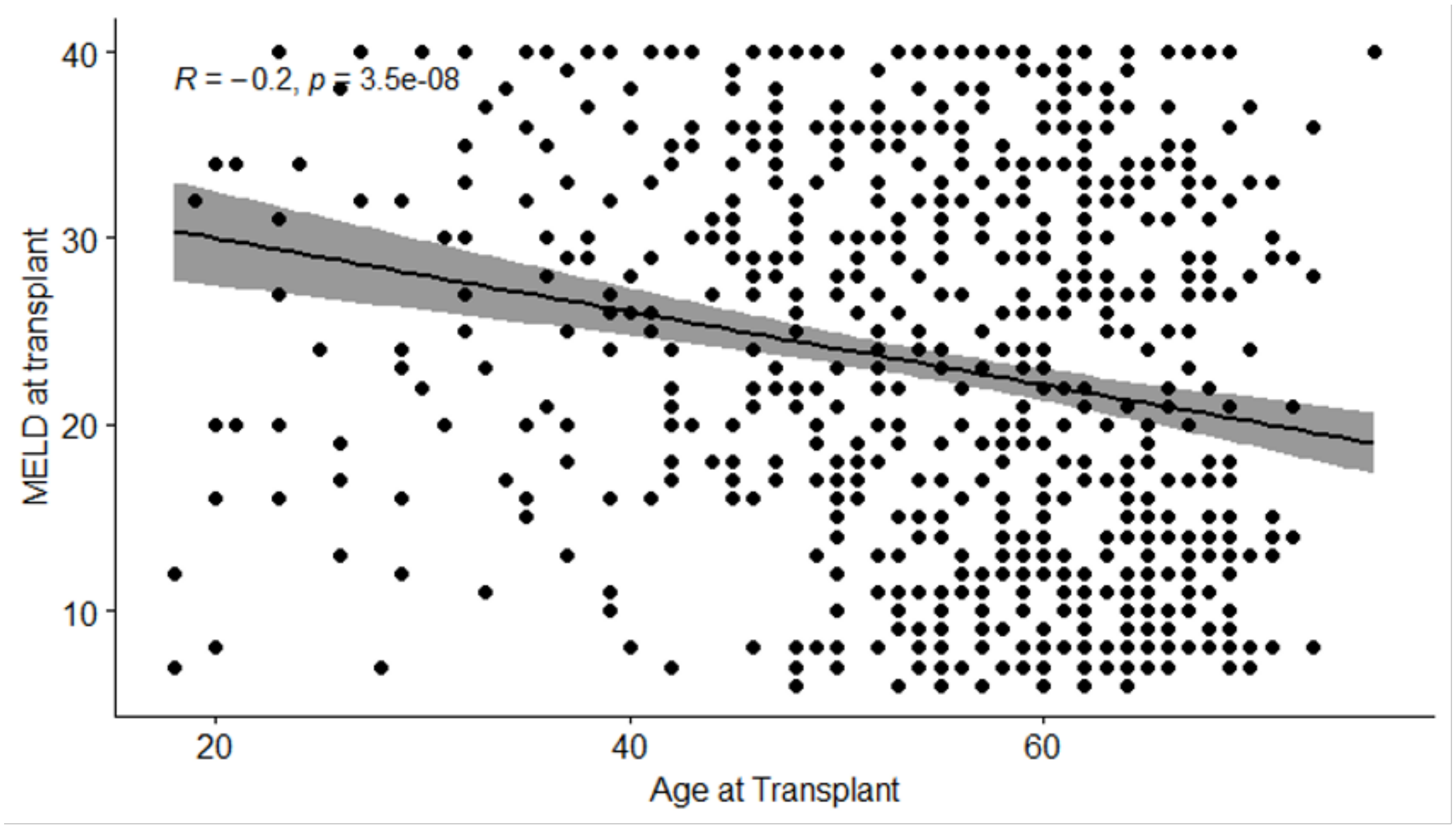

**Figure 1.** *Correlation between age and MELD at LT*

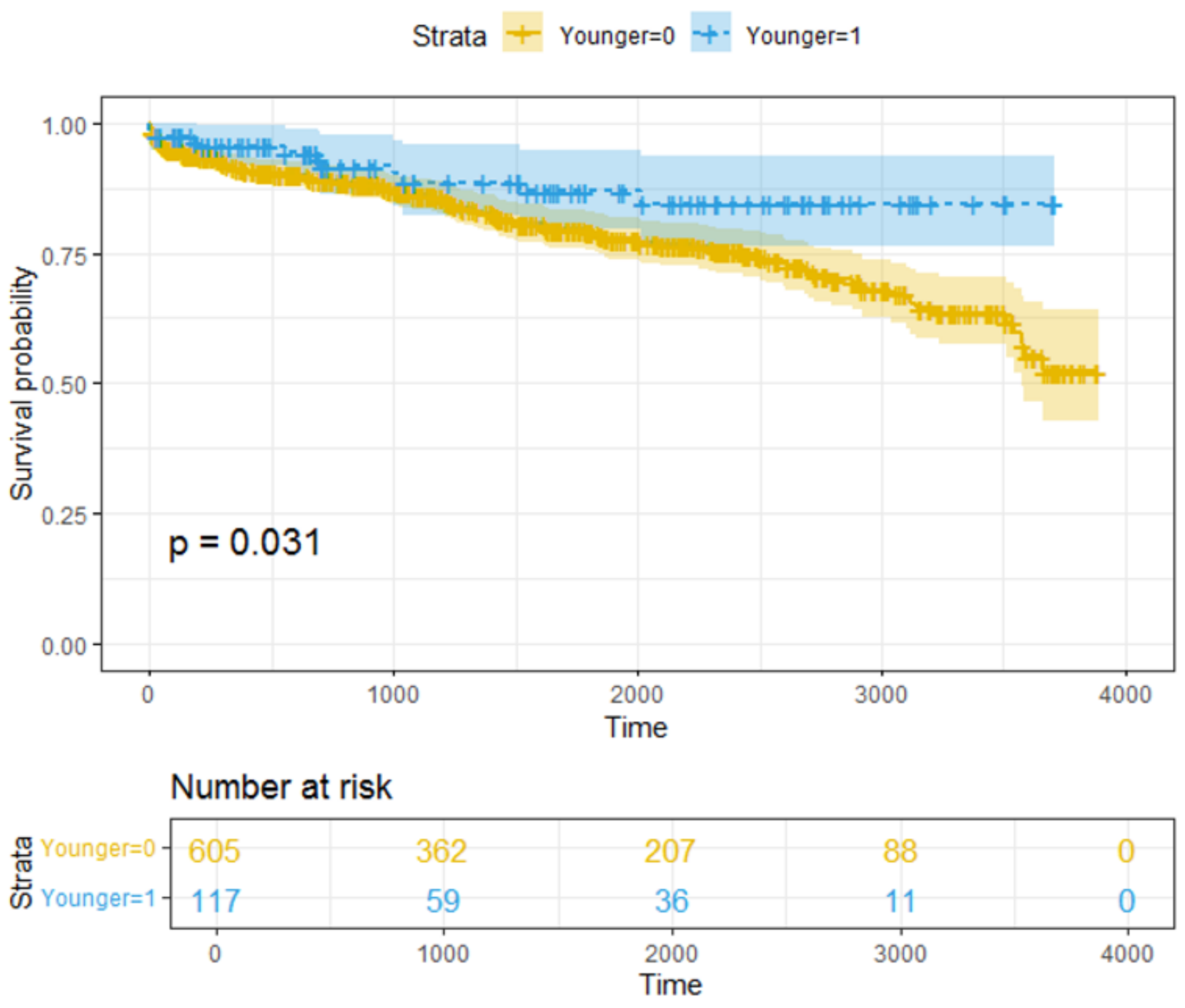

**Figure 2.** *Survival probabilities between patients with =< 45 years old and >45 years old at transplant.*

**Discussion**

This study aimed to determine risk factors impacting mortality post-LT, considering a multifactored strategy, developing an evidence- and data-driven approach capable of capturing the specific time ESLD comorbidities happens and whether the time when happens have an impact on mortality. To achieve that, first, the study identified the most frequent patients awaiting LT experience. Second, it used EBP guidelines to identify the diagnosis and staging metrics for each condition. Then finally, used the data extracted from the EBP guidelines to determine the time and occurrence of ESLD comorbidities in a real-word cohort of LT patients to analyze risk factors impacting mortality.

For this study and considering there is no validated LT-specific EBP for these population nor conditions, the most up to date EBP developed for the general population was used to extract evidence. However, a 2024 KDIGO updated guideline was published in March of 2024, one month following the closing of data extraction for this study. Upon review, diagnostic criteria do not differ greatly to change this study's results but should be further consider for any relevant changes that may have an impact on patient outcomes. Additionally, it is important to note that while all guidelines are accurate for use in the general population, ESLD population may have different thresholds for diagnosis. Further investigation should test whether and when comorbidities become harmful specifically for the LT population. Supplementary information to clarify and further capture the information needed for the matrices was taken from academic journals and accredited health organizations, as well as validated with content experts taking care of patients waiting for LT.

For a select few comorbidities, standard treatments currently in place in clinical practice, specifically for the population included in this study, were used as complementary diagnostic information to make retrospective cohort diagnosis and measuring of progression more accurate as EHR patient data lacked important testing information. For DM, there was no testing information differentiating T1DM from T2DM, likely because T1DM was diagnosed before patients were waitlisted for LT or outside the healthcare setting included in this study. Therefore, both T1DM from T2DM were considered as one singular condition for analysis, despite the matrices delineating the two. Diagnostic testing meeting the criteria for diagnosis, including the amount, type, and timing of each, was also limited in the patient data, so insulin use was measured as an additional variable to capture a more accurate set of data.

For CKD identification, although staging is based on both eGFR and albuminuria values, eGFR was the primary laboratory value available in this cohort. For this reason, patients considered in the early stages of CKD may be skewed as it lacks the inclusion of patients with a normal eGFR but do have persistent albuminuria. For malnutrition, EBP guidelines present many diagnostic criteria as it may not always be an objective finding. Non-laboratory testing data is the main source for diagnosis, such as dieticians' evaluation documented as free-text notes, and should be further validated using complementary laboratory tests, a thorough and accurate health history, and provider discretion. The inclusion of albumin,[16] a marker for malnutrition, was important to further quantify patients with malnutrition and capture the impact of nutritional status on post-LT survival. However, albumin is also absorbed and processed by the liver,[16] and a comprehensive understanding of malnutrition and its impact on ESLD patients are still to be elucidated. For portal hypertension,[14] the gold standard diagnostic tool is catheterization of the hepatic vein with the measure of portal pressure,[17] but it is an invasive, expensive, and scarce procedure, and the portal pressure result was not available, or because was documented in other formats in the EHR or because many clinicians do not document this precise of information.

Overall, the matrices were built to guide this study as a supplementary tool with condensed, quick, and easy-to-follow information to extract, identify and quantity the ESLD comorbidities. Thus, in this study, the reference EBP guidelines, even if primarily built for the general population, served as the primary guide to evaluate comorbidities in the LT cohort. The relevance of such an approach is to develop foundational knowledge that could be tested for efficacy in its clinical application, as well as to validate the occurrence of each comorbidity to mitigate provider variability in its diagnosis. While multiple organ system failure and health deterioration in ESLD patients are multifactorial in etiology, defining diagnostic guidelines using current EBP data could lead to earlier and more accurate identification that is standardized across clinicians and settings.

Establishing which conditions maintain statistical significance on LT outcomes was an important aspect of this study. Older age and longer time spent on the waitlist were consistently associated with a higher chance of mortality post-LT; a key finding that should play a role in decision-making. As shown by previous study, MELD scores at the time of transplant or at any point when used as a static metric, were not a consistently significant predictor of LT outcomes. This furthers the argument that MELD scores are a cross-section measure that does not capture disease variability over time; however, is still a good metric to predict 90-days mortality on the waitlist, but not post-LT.

On the other hand, DM at listing and developing low albumin and CKD stage 5 was found to increase the likelihood of mortality post-LT. As patients have different times in the waitlist, the challenge is still to determine when these findings become harmful for patients as many patients with these conditions recover well post-LT. As for both the survival and Cox proportional hazard tables, developing CKD stage 5 while on the waitlist was statistically significant, and better prediction tools should be developed to determine what are the ESLD patients who will progress to advanced CKD stages, what are the factors impacting this progression, and what treatments would be effective in changing these trajectories.

Considering the state of ESLD and the impact on nutritional absorption, malnutrition was not statistically significant in predicting LT outcomes in our studied cohort and was not included in the final model. However, low albumin was found significant,[18] and could have captured patients with malnutrition. Malnutrition is a major underlying condition impacting patient frailty, and research shows there is a highly significant association between frailty in ESLD patients and post-transplant mortality[19].[20] Perhaps it could be related that malnutrition is earlier diagnosed and, thus, improved management of these patients are implemented. It also may be that malnutrition data still lacks objectivity, and thus, our approach was not able to accurately identify all malnutrition cases nor the level of malnutrition.

Although insulin use has been shown to potentially impact patient outcomes, in this study, insulin was not a risk factor for mortality post-LT. However, DM at listing was a statistically significant predictor. It may be that patients with DM but in use of insulin are with the condition under control and, thus, not impacting mortality. This study findings confirm previous studies,[21] where DM and renal insufficiency were found to be significant predictors for worsened post-LT outcomes. Another retrospective cohort study[22] found that 56% of the cohort had underlying comorbidities with renal impairment being the most common; a similar finding to this study. Overall, both of these studies highlight that whole-health comorbidities are major factors in predicting survival outcomes following transplantation and should be more widely studied and considered in clinical settings.

Generally, this study faced challenges and some limitations that need to be considered. One of the limitations was the retrospective approach, where the analyzed data were reused and not collected primarily for the purpose of this study. Additionally, there was selection bias in the patient data as it was collected from a single-center institution with a primarily midwestern population, not accounting for population and geographic differences. Even considering all these challenges, strength exists. This study was able to identify the exact time of occurrence of comorbidities and use them to determine the cohort's likelihood of mortality after LT. A major strength of this study that facilitated this outcome included the use of time-specific data. EHR patient data is granular and provides detailed measures with second-specific time stamps. Furthermore, LT is an acute and complex procedure, so patients are followed very closely throughout the process by the transplant center, providing more information and data for accuracy.

**Conclusions**

A combination of EBP and data-driven research has the potential to improve the understanding of ESLD progression. With improved understanding, better management of the LT population can be implemented to improve outcomes so that challenges contributing to gaps in care can be better addressed. While factors such as a patient's age at LT cannot be changed, strategies addressing other important conditions can drive the future of personalized patient care in the ESLD population.

**Appendix**

Diabetes Mellitus Evidence-Based Practice Guideline Matrix

| **Prediabetes** | | | | | |
|---|---|---|---|---|---|
| **Description of Problem** | Glucose or Hemoglobin A1c levels do not meet the criteria for diabetes, yet have abnormal carbohydrate metabolism that results in elevated glucose levels (dysglycemia) intermediate between normoglycemia and diabetes | | | | |
| **Diagnosis of Staging** | **Validated Diagnostic Tools** | **Description of Validated Diagnostic Tools** | **Source** | **Minimum Threshold** | **Maximum Threshold** |
| Requires <u>two</u> *elevated* diagnostic laboratory test results measured either at the same time or at a second, prompt time point[a] | Hemoglobin A1c (HbgA1c) | Reflects a weighted average of glucose bound to hemoglobin over the lifespan of the erythrocyte (120 days) | Blood | 5.7% | 6.4% |
| | Fasting Plasma Glucose (FPG) | Measures serum glucose levels after fasting for at least an 8-hour period (no caloric intake) | Serum or Plasma | 100 mg/dL (5.6 mmol/L) | 125 mg/dL (6.9 mmol/L) |
| | 2-hour Glucose during 75-g Oral Glucose Tolerance Test (2-h PG; OGTT) | Measures serum glucose levels 2 hours after using a glucose load containing the equivalent of 75 g anhydrous glucose dissolved in water | Serum or Plasma | 140 mg/dL (7.8 mmol/L) | 199 mg/dL (11.0 mmol/L) |
| **Diabetes Mellitus Type 1** | | | | | |
| **Description of Problem** | Autoimmune b-cell destruction, usually leading to absolute insulin deficiency, including latent autoimmune diabetes in adults | | | | |
| **Diagnosis of Staging** | **Validated Diagnostic Tools** | **Description of Validated Diagnostic Tools** | **Source** | **Minimum Threshold** | **Maximum Threshold** |
| Requires <u>either</u> classic symptoms of hyperglycemia *and* a random plasma glucose of ≥200 mg/dL [≥11.1 mmol/L] <u>or</u> *two* abnormal diagnostic laboratory test results measured either at the same time or at a second, prompt time point[a] with evidence of autoimmune pancreatic islet cell damage | Pancreatic Islet Cell Antibodies (ICAs) OR Anti-insulin autoantibodies (AI-2) | Screens for the presence of autoantibody proteins in the pancreatic islet which indicates T1DM from other forms | Serum | Positive | |
| | Hemoglobin A1c (HbgA1c) | Reflects a weighted average of glucose bound to hemoglobin over the life span of the erythrocyte (120 days) | Blood | 6.5% | n… |
| | Fasting Plasma Glucose (FPG) | Measures serum glucose levels after fasting for at least an 8-hour period (no caloric intake) | Serum or Plasma | 126 mg/dL (7.0 mmol/L) | n… |
| | 2-hour Glucose during 75-g Oral Glucose Tolerance Test (2-h PG; OGTT) | Measures serum glucose levels 2 hours after using a glucose load containing the equivalent of 75 g anhydrous glucose dissolved in water | Serum or Plasma | 200 mg/dL (11.1 mmol/L) | n… |
| | Random Blood Glucose (RBG) occurring with hyperglycemic symptoms | Measures plasma glucose without regard to time since previous meal with the occurrence of hyperglycemic symptoms (i.e. polyuria, polydipsia, and unexplained weight loss) | Blood | 200 mg/dL (11.1 mmol/L) | n… |
| | Clinical Findings (H&P) | Assessment for the presence of the three “Ps”: polyuria, polydipsia, polyphagia | Complementary: Signs & Symptoms (S&S) | Present | |
| | Medication Use (H&P) | Assessment for current or long-term use of insulin medication to manage blood glucose levels regularly | Complementary: Medication Reconciliation | Present | |
| **Severity Measures** | Increasing insulin medication needs until complete insulin dependence for blood glucose regulation | | | | |

| | |
|---|---|
| | Increasing HbgA1c levels 120 days after the most recent result |
| | Incidence of diabetic complications |
| | Hospital stays related to diabetic or glucose dysregulation |
| **Complications** | **Acuity** |
| Diabetic Ketoacidosis (DKA) | Acute |
| Hypoglycemia | Acute |
| Nephropathy | Chronic |
| Neuropathy | Chronic |
| Retinopathy | Chronic |
| Infection | Acute |
| Hypertension | Chronic |
| **Diabetes Mellitus Type 2** | |
| **Description of Problem** | Non-autoimmune progressive loss of adequate b-cell insulin secretion, frequently on the background of insulin resistance and metabolic syndrome |

| **Diagnosis of Staging** | **Validated Diagnostic Tools** | **Description of Validated Diagnostic Tools** | **Source** | **Minimum Threshold** | **Maximum Threshold** |
|---|---|---|---|---|---|
| Requires <u>either</u> classic symptoms of hyperglycemia *and* a random plasma glucose of ≥200 mg/dL [≥11.1 mmol/L] <u>or</u> *two* abnormal diagnostic laboratory test results measured either at the same time or at a second, prompt time point[a] | Hemoglobin A1c (HbgA1c) | Reflects a weighted average of glucose bound to hemoglobin over the life span of the erythrocyte (120 days) | Blood | 6.5% | n… |
| | Fasting Plasma Glucose (FPG) | Measures serum glucose levels after fasting for at least an 8-hour period (no caloric intake) | Serum or Plasma | 126 mg/dL (7.0 mmol/L) | n… |
| | 2-hour Glucose during 75-g Oral Glucose Tolerance Test (2-h PG; OGTT) | Measures serum glucose levels 2 hours after using a glucose load containing the equivalent of 75 g anhydrous glucose dissolved in water | Serum or Plasma | 200 mg/dL (11.1 mmol/L) | n… |
| | Random Blood Glucose (RBG) occurring with hyperglycemic symptoms | Measures plasma glucose without regard to time since previous meal with the occurrence of hyperglycemic symptoms (i.e. polyuria, polydipsia, and unexplained weight loss) | Blood | 200 mg/dL (11.1 mmol/L) | n… |
| | Clinical Findings (H&P) | Assessment for the presence of the three "Ps": polyuria, polydipsia, polyphagia | Complementary: Signs & Symptoms (S&S) | Present | |
| | Medication Use (H&P) | Assessment for current or long-term use of oral antidiabetics or insulin medication to manage blood glucose levels regularly | Complementary: Medication Reconciliation | Present | |
| **Severity Measures** | Increasing the number or dosage of oral antidiabetic medications | | | | |

| | |
|---|---|
| | Increasing from oral antidiabetic medications to total dependence on insulin replacement for blood glucose regulation |
| | Increasing HbgA1c levels 120 days after the most recent result |
| | Incidence of diabetic complications |
| | Hospital stays related to diabetic or glucose dysregulation |
| **Complications** | **Acuity** |
| Diabetic Ketoacidosis (DKA) | Acute |
| Hypoglycemia | Acute |
| Nephropathy | Chronic |
| Neuropathy | Chronic |
| Retinopathy | Chronic |
| Infection | Acute |
| Cardiovascular Disease (CVD) | Chronic |
| Peripheral vascular disease (PVD) | Chronic |
| Stroke | Acute |
| Myocardial Infarction | Acute |
| Chronic kidney disease | Chronic |

*Note*. Data are from (American Diabetes Association, n.d.; American Diabetes Association Professional Practice Committee, 2023; Kahanovitz et al., 2017; Orban et al., 2009)

[a]Can be a different test or a repeat test

Chronic Kidney Disease Evidence-Based Practice Guideline Matrix

| **Chronic Kidney Disease** | |
|---|---|
| **Description of Problem** | Abnormalities of kidney structure or function, present for more than 3 months, with implications for health |
| **Diagnosis** | Requires either a decreased Glomerular Filtration Rate (eGFR)[a] below 60 ml/min/1.73 $m^2$ present for more than 3 months or one or more markers for kidney damage (See Below) |
| | Albuminuria classified by an Albumin-to-Creatinine Ratio (ACR) >30 mg/g |
| | Urine sediment abnormalities |
| | Electrolyte and other abnormalities due to tubular disorders |

| | | | | | | |
|---|---|---|---|---|---|---|
| | Abnormalities detected by histology | | | | | |
| | Structural abnormalities detected by imaging | | | | | |
| | History of kidney transplantation | | | | | |
| **Staging** | **Description of Staging** | | | | | |
| Stage 1 | Normal or high eGFR (G1) with persistent albuminuria (A1, A2, or A3) | | | | | |
| Stage 2 | Mildly decreased eGFR (G2) with persistent albuminuria (A2 or A3) | | | | | |
| Stage 3a | Mildly to severely decreased eGFR (G3a or 3b) with or without albuminuria | | | | | |
| Stage 3b | Mildly to severely decreased eGFR (G3a or 3b) with or without albuminuria | | | | | |
| Stage 4 | Severely decreased eGFR (G4) with or without albuminuria | | | | | |
| Stage 5 | Kidney failure eGFR level (G5), irrespective of albuminuria | | | | | |
| **Categories** | **Description of Categories** | **Validated Diagnostic Tools** | **Description of Validated Diagnostic Tools** | **Source** | **Minimum Threshold** | **Maximum Threshold** |
| eGFR (See Below) | | **Glomerular Filtration Rate (eGFR)** | Measures the reduction in kidney excretory function that is generally reduced after widespread structural damage | Serum, Plasma, or Blood | | |
| G1 | Normal or high | | | | 90 ml/min/1.73 m$^2$ | n… |
| G2 | Mildly decreased | | | | 60 ml/min/1.73 m$^2$ | 89 ml/min/1.73 m$^2$ |
| G3a | Mildly to moderately decreased | | | | 45 ml/min/1.73m$^2$ | 59 ml/min/1.73 m$^2$ |
| G3b | Moderately to severely decreased | | | | 30 ml/min/1.73 m$^2$ | 44 ml/min/1.73 m$^2$ |
| G4 | Severely decreased | | | | 16 ml/min/1.73 m$^2$ | 29 ml/min/1.73 m$^2$ |
| G5 | Kidney Failure | | | | 0 ml/min/1.73 m$^2$ | 15 ml/min/1.73 m$^2$ |
| Albuminuria- ACR (See Below) | | Albumin-to-Creatinine Ratio (ACR) | Measures abnormal loss of plasma protein (albumin) in the urine | Urine | | |
| A1 | Normal to mildly increased | | | | 0 mg/g | 30 mg/g |
| A2 | Moderately increased | | | | 30 mg/g | 300 mg/g |
| A3 | Severely increased | | | | 300 mg/g | n… |
| **Severity Measures** | Drop in the eGFR staging category accompanied by a 25% or greater drop in eGFR from baseline (small fluctuations in eGFR are common and not indicative of disease progression) | | | | | |
| | Rapid progression is defined as a sustained decline in eGFR of more than 5 ml/min/1.73 m$^2$/yr | | | | | |
| | Lower eGFR and greater albuminuria are both associated with an increased rate of progression (synergistic; Patients with a lower eGFR and greater albuminuria are more likely to progress) | | | | | |
| | Hospital stays related to kidney disease progression or acute kidney injuries | | | | | |

| | |
|---|---|
| | Incidence of CKD complications |
| | Initiation of renal replacement therapy or dialysis |
| **Complications** | **Acuity** |
| Acute Kidney Injury | Acute |
| Drug Toxicity | Acute |
| Cardiovascular Disease (CVD) | Chronic |
| Infection | Acute |
| Frailty | Chronic |
| Cognitive Impairment | Acute or Chronic |
| Endocrine and Metabolic Conditions | Chronic |
| Anemia | Chronic |
| Electrolyte and Fluid Imbalances | Acute |

*Note*. Data are from (Cleveland Clinic, n.d.-b; Kidney Disease Improving Global Outcomes (KDIGO), 2013)

[a]Best overall index of kidney function in health and disease

Malnutrition Evidence-Based Practice Guideline Matrix

<table>
<tr><td colspan="6">Malnutrition</td></tr>
<tr><td>Description of Problem</td><td colspan="5">State resulting from lack of uptake or intake of nutrition leading to altered body composition (decreased fat-free mass) and body cell mass leading to diminished physical and mental function and impaired clinical outcome from disease</td></tr>
<tr><td colspan="6">Diagnosis</td></tr>
<tr><td rowspan="3">Requires either Alternative A or Alternative B</td><td>Alternative A</td><td colspan="4">Underweight Body Mass Index (BMI) classified as ≤18.5 kg/m$^2$</td></tr>
<tr><td rowspan="2">Alternative B</td><td rowspan="2">Unintentional weight loss (wt. %) combined with either:</td><td colspan="3">An abnormal decrease in BMI[a] dependent on age or</td></tr>
<tr><td colspan="3">An abnormal decrease in Fat-Free Mass Index (FFMI) dependent on gender</td></tr>
<tr><td rowspan="4">Complementary Diagnostic Measures</td><td>Validated Diagnostic Tools</td><td>Description of Validated Diagnostic Tools</td><td>Source</td><td>Minimum Threshold</td><td>Maximum Threshold</td></tr>
<tr><td>Albumin[a]</td><td>Hepatic protein with a half-life of 14-20 days; functions as a carrier hormone for many minerals, hormones, and fatty acids</td><td>Serum</td><td>0 g/dL</td><td>3.3 g/dL</td></tr>
<tr><td>Prealbumin (PAB)[b]</td><td>Hepatic, negative, acute-phase protein with a half-life of 2-3 days</td><td>Serum</td><td>0 mg/dL</td><td>16 mg/dL</td></tr>
<tr><td>Nutrition Therapy (H&P)</td><td>Assessment for current or long-term use of standard oral nutrition supplements and/or enteral or parenteral nutrition formulas when caloric and protein intake from a regular diet is not adequate to meet nutritional needs</td><td>Assessment</td><td>Present</td><td></td></tr>
</table>

| Staging | Diagnosis of Staging | Validated Diagnostic Tools | Description of Validated Diagnostic Tools | Source | Minimum Threshold | Maximum Threshold |
|---|---|---|---|---|---|---|
| Stage 1- Moderate | Requires *one* abnormal diagnostic test result in either Weight Loss %, Body Mass Index, or Muscle Mass | Weight Loss (wt. %) | Phenotypic criterion tool used to assess pacing of body weight loss over time to identify trajectories of decline, maintenance, and improvement | Minimum Data Set | Within 6 months: 5% | Within 6 months: 10% |
| | | | | | Beyond 6 months: 10% | Beyond 6 months: 20% |
| | | Body Mass Index (BMI) | Phenotypic criterion tool used as an indicator of body density as determined by the relationship of body weight to body height | Ratio | <70 years old: 0 $kg/m^2$ | <70 years old: 20 $kg/m^2$ |
| | | | | | ≥70 years old: 0 $kg/m^2$ | ≥70 years old: 22 $kg/m^2$ |
| | | Body Muscle Mass (MM) (Per validated assessment tool- See Below) | Phenotypic criterion tool that uses many different validated assessment methods to measure muscle mass loss based on availability in clinical setting | Core Component or Analyte Measured | Mild deficit | Moderate deficit |
| | | Fat-Free Mass Index (FFMI) | Measures mainly muscle mass, or any fat-free body mass | Ratio | Female: 0 $kg/m^2$ | Female: 15 $kg/m^2$ |
| | | | | | Male: 0 $kg/m^2$ | Male: 17 $kg/m^2$ |
| | | Appendicular Skeletal Muscle index (ASMI) | Calculates the muscle mass of the 4 limbs | Ratio | | |
| | | Dual-Energy X-ray Absorptiometry (DXA) | Estimates the total amount of lean tissue but does not directly measure muscle mass | Medical Imaging | | |
| | | Bioelectrical Impedance Analysis (BIA) | Measures body composition using different bioimpedance measurement techniques | Variety | | |
| | | Appendicular Lean Mass (ALM) | Measures the sum of the lean tissue in the arms and legs derived from DXA scans | Variety | | |
| | | Calf or Mid-Arm Circumference (CC; MAC) | Measures circumference of muscle area | Measurement | | |
| | | Hand Grip Strength (HGS) | Measures maximum voluntary muscle strength | Measurement | | |
| Stage 2-Severe | Requires *one* abnormal diagnostic test result in <u>either</u> Weight Loss %, Body Mass Index, <u>or</u> Muscle Mass | Weight (wt. %) | Phenotypic criterion tool used to assess pacing of body weight loss over time to identify trajectories of decline, maintenance, and improvement | Minimum Data Set | Within 6 months: 10% | Within 6 months: n… |
| | | | | | Beyond 6 months: 20% | Beyond 6 months: n… |
| | | Body Mass Index (BMI) | Phenotypic criterion tool used as an indicator of body density as determined by the relationship of <u>body weight</u> to <u>body height</u> | Ratio | <70 years old: 0 $kg/m^2$ | <70 years old: 18.5 $kg/m^2$ |
| | | | | | ≥70 years old: 0 $kg/m^2$ | ≥70 years old: 20 $kg/m^2$ |
| | | Body Muscle Mass (MM) (Per validated assessment tool- See Below) | Phenotypic criterion tool that uses many different validated assessment methods to measure muscle mass loss based on availability in the clinical setting | Core component or analyte measured | Severe deficit | |
| | | Fat-Free Mass Index (FFMI) | Measures mainly muscle mass, or any fat-free body mass | Ratio | Female: 0 $kg/m^2$ | Female: 15 $kg/m^2$ |
| | | | | | Male: 0 $kg/m^2$ | Male: 17 $kg/m^2$ |
| | | Appendicular Skeletal Muscle index (ASMI) | Calculates the muscle mass of the 4 limbs | Ratio | | |
| | | Dual-Energy X-ray Absorptiometry (DXA) | Estimates the total amount of lean tissue but does not directly measure muscle mass | Medical Imaging | | |
| | | Bioelectrical Impedance Analysis (BIA) | Measures body composition using different bioimpedance measurement techniques | Variety | | |
| | | Appendicular Lean Mass (ALM) | Measures the sum of the lean tissue in the arms and legs derived from DXA scans | Variety | | |
| | | Calf or Mid-Arm Circumference (CC; MAC) | Measures circumference of muscle area | Measurement | | |
| | | Hand Grip Strength (HGS) | Measures maximum voluntary muscle strength | Measurement | | |

| | |
|---|---|
| **Severity Measures** | Decreasing BMI |
| | Increasing unintentional weight loss |
| | Loss of muscle mass or muscle wasting |
| | Increasing number or dosage of supplemental nutrition therapy |
| | Incidence of malnutrition complications |
| | Hospital stays related to deterioration in nutritional status or related complications |
| **Complications** | **Acuity** |
| Hypoglycemia | Acute |
| Hypothermia | Acute |
| Dehydration | Acute |
| Electrolyte Imbalance | Acute |
| Diarrhea | Acute or Chronic |
| Frailty | Chronic |
| Loss of Functional Capacity | Chronic |
| Refeeding Syndrome | Acute |
| Decrease in cardiac output | Chronic |
| Acute Kidney Injury | Acute |
| Respiratory Tract Infection | Acute |
| Infection | Acute |
| Delayed wound healing | Acute |
| Psychosocial effects | Chronic |

*Note*. Data are from (Bharadwaj et al., n.d.; Cawthon, 2015; Cederholm et al., 2015, 2019; Keller, 2019; Khalil et al., 2014; Lai et al., 2021; Plauth et al., n.d., 2006, 2019; Saunders & Smith, 2010)

[a]Albumin levels are highly altered during many inflammatory and disease processes, therefore it is an unreliable marker for malnutrition when taken alone

[b]Prealbumin (PAB) has a much shorter half-life than albumin, so is more reliable for measuring acute changes in nutrition states. However, PAB levels are also affected by inflammatory states therefore is also unreliable in diagnosing malnutrition when taken alone. PAB is also affected by renal and thyroid dysfunction as the kidneys degrade it and it serves as a transport protein for thyroxine.

Portal Hypertension Evidence-Based Practice Guideline Matrix

<table>
<tr><th colspan="6">Portal Hypertension</th></tr>
<tr><td>Description of Problem</td><td colspan="5">A portal pressure gradient of the difference in pressure between the portal vein and the hepatic veins, >5 mm Hg</td></tr>
<tr><td colspan="6">Diagnosis</td></tr>
<tr><td rowspan="6">Requires <u>either</u> one or more invasive tool finding <u>or</u> non-invasive tool finding</td><td rowspan="3">Invasive Tool Findings</td><td colspan="4">Elevated hepatic venous pressure gradient >5 mm Hg: GOLD STANDARD</td></tr>
<tr><td colspan="4">Gastroesophageal varices</td></tr>
<tr><td colspan="4">Past placement of a Transjugular Intrahepatic Portosystemic Shunt (TIPS)</td></tr>
<tr><td rowspan="3">Noninvasive Tool Findings</td><td colspan="4">Increased liver stiffness</td></tr>
<tr><td colspan="4">Increased liver stiffness, decreased platelet count, <u>and</u> splenomegaly</td></tr>
<tr><td colspan="4">Portosystemic collaterals (shunts) <u>and/or</u> reversal of flow within the portal system</td></tr>
<tr><td>Staging</td><td>Validated Diagnostic Tools</td><td>Description of Validated Diagnostic Tools</td><td>Source</td><td>Minimum Threshold</td><td>Maximum Threshold</td></tr>
<tr><td>Mild; Compensated</td><td rowspan="2">Catheterization of the hepatic vein[a]: GOLD STANDARD</td><td rowspan="2">Determines the hepatic venous pressure gradient, which is the difference between the wedged (or occluded) hepatic venous pressure and the free hepatic venous pressure using a balloon catheter</td><td rowspan="2">Invasive Tool: Catheterizatio n</td><td>5 mm Hg</td><td>10 mm Hg</td></tr>
<tr><td>Clinically Significant; Decompensated</td><td>10 mm Hg</td><td>n…</td></tr>
</table>

| | | | | | |
|---|---|---|---|---|---|
| | Esophagogastroduodenoscopy (EGD) | Visualizes for the presence of gastroesophageal varices using an endoscope | Invasive Tool: Endoscopy | Present | |
| | Past placement of a Transjugular Intrahepatic Portosystemic Shunt (TIPS) | A shunt is used to connect the portal vein to adjacent hepatic veins to bypass the pressure within the veins of the liver | Invasive Tool: procedure | Past Presence | |
| | Transient elastography; FibroScan (TE) | Assesses the stiffness of the liver, a physical property of liver tissue influenced by the amount of liver fibrosis content, using an ultrasound transducer | Noninvasive Tool: Medical Imaging | 20 kPa | 25 kPa |
| | Platelet count (PLT) | Measures number of platelet fragments which plays a major role in the cascade of events that lead to clot formation | Noninvasive Tool: Platelets | 0 platelets/uL | 150,000 platelets/uL |
| | Ultrasound of Spleen (US) | Uses sound waves to produce images of the spleen to measure its size | Noninvasive Tool: Medical Imaging | 13 cm | n… |
| | Computed Tomography of Spleen (CT) | Uses a combination of X-rays and computer technology to produce images of the spleen to measure its size | Noninvasive Tool: Medical Imaging | 13 cm | n… |
| | Magnetic Resonance Imaging of Spleen (MRI) | Uses strong magnetic fields, magnetic field gradients, and radio waves to produce images of the spleen to measure its size | Noninvasive Tool: Medical Imaging | 13 cm | n… |
| | Ultrasound of Abdomen and Liver (US) | Uses sound waves to produce images of the abdominal and hepatic regions for the presence of portosystemic collaterals (shunts) and/or reversal of flow within the portal system | Noninvasive Tool: Medical Imaging | Present | |
| | Computed Tomography of Abdomen and Liver (CT) | Uses a combination of X-rays and computer technology to produce images of the abdominal and hepatic regions for the presence of portosystemic collaterals (shunts) and/or reversal of flow within the portal system | Noninvasive Tool: Medical Imaging | Present | |
| | Magnetic Resonance Imaging of Abdomen and Liver (MRI) | Uses strong magnetic fields, magnetic field gradients, and radio waves to produce images of the abdominal and hepatic regions for the presence of portosystemic collaterals (shunts) and/or reversal of flow within the portal system | Noninvasive Tool: Medical Imaging | Present | |
| | Physical Examination (H&P) | Physical assessment for the presence of spider nevi and/or visible abdominal portosystemic collaterals, but its absence cannot be used to rule out clinically significant portal hypertension therefore it is a complementary tool | Complementary Noninvasive Tool: Inspection | Present | |
| **Severity Measures** | | Increasing hepatic venous pressure gradient from mild to clinically significant portal hypertension (and beyond) | | | |
| | | Increasing number of formation and growth of gastroesophageal varices | | | |
| | | Progressive stiffening of liver tissue | | | |
| | | Decreasing platelet count | | | |
| | | Progressive spleen enlargement | | | |
| | | Appearance of new portosystemic collaterals (shunts) | | | |
| | | Incidence of portal hypertension complications or decompensating events | | | |
| | | Hospital stays related to increasing severity of portal hypertension or related complications | | | |

| Complications | Acuity |
|---|---|
| Hemorrhage | Acute |
| Ascites | Acute or Chronic |
| Hepatic Encephalopathy | Acute or Chronic |
| Gastroesophageal Varices | Chronic |
| Gastroesophageal Varices Hemorrhage | Acute |

*Note*. Data are from (Cleveland Clinic, n.d.-a; Garcia-Tsao et al., 2017; Mount Sinai, 2024)

[a]Limitations of Catheterization procedure: invasive, expensive, limited availability

Ascites Evidence-Based Practice Guideline Matrix

| **Ascites** | | | | |
|---|---|---|---|---|
| **Description of Problem** | Abnormal buildup of fluid in the peritoneal cavity of the abdomen | | | |
| **Diagnosis** | Requires (either) fluid confirmation by abdominal ultrasound, and (or) a successful abdominal paracentesis | | | |
| **Validated Diagnostic Tools** | **Description of Validated Diagnostic Tools** | **Source** | **Minimum Threshold** | **Maximum Threshold** |
| Ultrasound of Abdomen (US) | Uses sound waves to produce images of the abdominal region to determine with certainty the presence of fluid in the peritoneal cavity of the abdomen | Medical Imaging | Present | |
| Abdominal Paracentesis | Insertion of a needle into the left lower quadrant of the abdominal wall to aspirate and remove ascitic fluid from the peritoneal cavity | Procedure | 0 mL | n… |
| Ascitic Fluid Analysis (See Below) | Assesses the abnormal collection of fluid withdrawn from the peritoneal cavity during the abdominal paracentesis for diagnosis of its etiology | Peritoneal Fluid | | |
| Cell Count and Differential | Assesses for the presence of cells in the aspirated peritoneal fluid | Peritoneal Fluid | Present | |
| Serum-Ascites Albumin Gradient (SAAG) | Determines the mass concentration difference of albumin in the serum versus the peritoneal fluid and the cause of ascites; SAAG = (Albumin concentration of serum) - (Albumin concentration of ascitic [peritoneal] fluid) | Calculation: Peritoneal Fluid and Serum | Caused by Portal Hypertension: 1.1 g/dL<br><br>Not Caused by Portal Hypertension: 0 g/dL | Caused by Portal Hypertension: n…<br><br>Not Caused by Portal Hypertension: 1.1 g/dL |

<table>
<tr><td>Total Protein (TP)</td><td>Assesses for the presence of protein in the aspirated peritoneal fluid</td><td>Peritoneal Fluid</td><td>Present</td><td></td></tr>
<tr><td>Physical Examination (H&P)</td><td>Physical assessment technique using the hands as an instrument to percuss on the lateral aspect of the abdomen with the patient in a supine position for a greater dullness (sound) than usual indicating the presence of fluid in the abdominal cavity and shifting dullness</td><td>Complementary: Percussion</td><td>Present</td><td></td></tr>
<tr><td rowspan="5">Severity Measures</td><td colspan="4">Increasing or high amounts of an abnormal buildup of fluid in the peritoneal cavity of the abdomen</td></tr>
<tr><td colspan="4">Increasing dosage or the number of diuretic medication treatment therapies</td></tr>
<tr><td colspan="4">Inability to effectively manage condition noninvasively with diuretics, dietary sodium restriction, or others leading to invasive treatment options including the performance of an abdominal paracentesis</td></tr>
<tr><td colspan="4">Incidence of ascites complications</td></tr>
<tr><td colspan="4">Hospital stays related to increasing severity of ascites or related complications</td></tr>
<tr><td>Complications</td><td colspan="4">Acuity</td></tr>
<tr><td>Spontaneous Bacterial Peritonitis</td><td colspan="4">Acute</td></tr>
<tr><td>Hepatorenal Syndrome (Kidney Failure)</td><td colspan="4">Acute or Chronic</td></tr>
<tr><td>Malnutrition</td><td colspan="4">Chronic</td></tr>
<tr><td>Hepatic Encephalopathy</td><td colspan="4">Acute or Chronic</td></tr>
</table>

| | |
|---|---|
| Gastrointestinal Bleeding | Acute or Chronic |
| Pleural Effusion | Acute |

*Note*. Data are from (Penn Medicine, 2024; Runyon, 2009)